\documentclass[prb,preprint,superscriptaddress,showpacs,preprintnumbers,amsmath,amssymb]{revtex4}
\usepackage{graphicx}
\usepackage{dcolumn}
\usepackage{bm}
\begin{document}

\title{Coulomb pseudogap in elastic 2D-2D electron tunneling in a quantizing magnetic field}

\author{V. G. Popov}
\affiliation{Institute of Microelectonics Technology RAS,
Chernogolovka 142432, Russia}
\author{O. N. Makarovskii}
\affiliation{School of Physics and Astronomy of Nottingham
University, Nottingham NG7 2RD, UK}
\author{V. Renard}
\affiliation{GHMFL, CNRS, BP 166, F-38042, Grenoble Cedex 9,
France}
\author{L. Eaves}
\affiliation{School of Physics and Astronomy of Nottingham
University, Nottingham NG7 2RD, UK}
\author{J.-C. Portal}
\affiliation{GHMFL, CNRS, BP 166, F-38042, Grenoble Cedex 9,
France} \affiliation{INSA, 135 Avenue de Rangueil, F-31077,
Toulouse Cedex 4, France} \affiliation{Institut Universitaire de
France, 103 Bd. St. Michel, F-75005, Paris, France}
\date{\today}

\begin{abstract}
The electron tunneling is experimentally studied between
two-dimensional electron gases (2DEGs) formed in a
single-doped-barrier heterostructure in the magnetic fields
directed perpendicular to the 2DEGs planes. It is well known that
the quantizing magnetic field induces the Coulomb pseudogap
suppressing the electron tunneling at Fermi level. In this paper
we firstly present the experimental results revealing the
pseudogap in the electron tunneling assisted by elastic electron
scattering on disorder.
\end{abstract}

\maketitle

 In the tunnel structures with the two-dimensional electron gas (2DEG)
one of the first observed many-body phenomenon was the Coulomb
pseudogap\cite{Eisenstein}. The pseudogap means suppression of
tunneling at the Fermi level of the 2DES while the electron
transport is allowed in the lateral direction. Since the tunneling
is a very fast event it introduces a local charge fluctuation that
relaxes via the lateral dissipative current. In magnetic field
these current organizes a vortex that costs some finite
energy\cite{vortex}. Therefore the tunneling electron should have
some extra energy to organize its relaxation. The pseudogap was
observed as a suppression of the tunneling current at zero voltage
in the tunnel junction between the identical
2DEGs\cite{Eisenstein} or as an additional voltage shift of the
resonant current peak in the case of the different
2DEGs\cite{Geim}. In the magnetic field new current peaks appear
in the I-V curve those originate from the electron tunneling
assisted by elastic scattering. In this case the energy
conservation gives an additional resonant condition as follows:
\begin{equation}\label{elastic}
    E_{\rm{01}}-E_{\rm{02}} = k\hbar\omega_c.
\end{equation}
where $k$ is an integer number, $\hbar$ is the Planck constant,
$\omega_c$ is the cyclotron frequency. In other words the
scattering assists the electron tunneling between LLs with the
different indexes and k is the difference of these indexes. As it
was mentioned above the many-body effects cause the voltage shift
of the resonance peak. The question is what happens with the
elastic features? In this paper we present the results of the
experimental study of the elastic features in the 2D-2D electron
tunneling in the high magnetic fields. In particularly we have
first investigated situation when the cyclotron energy is higher
than the difference between the subband levels in the emitter and
collector 2DEGs or intersubband energy.

\begin{figure}
\includegraphics{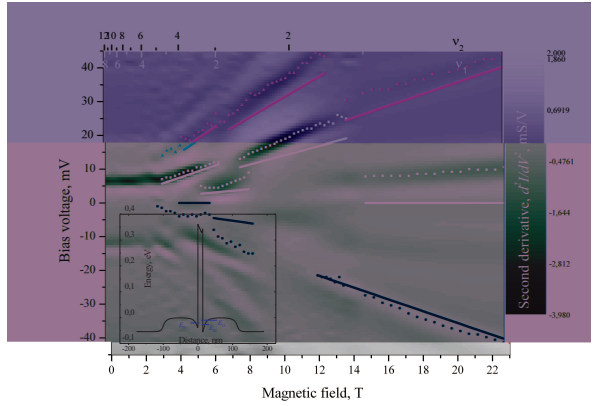} \caption{\label{fig:spectra} Tunnel spectra of the studied heterostructure
in the magnetic field applied perpendicular to the 2DEG planes.
Square symbols show $V_r$ values and lengths of the solid lines
shows the calculated ones.}
\end{figure}

The investigated tunnel diodes were made of a single-barrier
heterostructure by the conventional photolithography and the
wet-etching technique. The heterostructure was grown by the
molecular-beam epitaxy on a n$^+$-Si-doped GaAs substrate and
consists of modulated doped GaAs layers and a single 20~nm thick
layer of Al$_{0.3}$Ga$_{0.7}$As which is doped in the middle. The
schematic conduction-band-bottom diagram is shown in the insert in
Fig.~\ref{fig:spectra} with the quantum subband levels in the
2DEGs. The parameters of the 2DEGs are the following: the
concentration of the 2DEG with the level $E_{\rm{01}}$ is $n_1 =
4\times10^{11}$~cm$^{-2}$; the concentration of the 2DEG with
level $E_{\rm{02}}$ is $n_2 = 6 \times 10^{11}$~cm$^{-2}$. The
tunnel characteristics were measured at liquid $^3$He temperature
$T = 1.5$~K. Current peak or the second derivative minimum
corresponds to the first coherent resonance at the bias voltage
$V_r = 7$~mV, i. e., $E_{\rm{01}}(V_r) = E_{\rm{02}}(V_r)$. The
second derivative maximum at $V_s = - 14$~mV associates with the
second resonance, i. e., $E_{\rm{01}}(V_s) = E_{\rm{12}}(V_s)$. In
the magnetic field new features appear in the I-V curve and in the
second derivative (see Fig.~\ref{fig:spectra}).

To understand the non-monotonic voltage shift of the current peak
it is necessary to keep in mind that due to the charge transfer
between the 2DEGs and contacts the subband levels become
oscillating on the field. Moreover these oscillations can be quite
strong to provide LL pinning on Fermi levels. The oscillation
period is determined by the average value of the Fermi energy.
Since the concentrations of the 2DEGs are different the
oscillations of the subband levels $E_{\rm{01}}$ and $E_{\rm{02}}$
are not in-phase that causes the oscillations of the resonance
voltage $V_r$. The experimental values of the voltage $V_r$ are
plotted in Fig.~\ref{fig:spectra} as the square symbols. Similar
dependence was observed and had been successfully described
previously in Ref.~\onlinecite{pseudogap}. Following this model it
is worth noting that the coherent resonance takes place when the
ladders of the LLs coincide in the both 2DEGs. If the upper LLs
are pined to the Fermi levels in both 2DEGs then the resonance
voltage will be determined by the energy difference of the pined
upper LLs. One can define the upper LLs from the LL filling
factors plotted in the top axes in Fig.~\ref{fig:spectra}. In
accordance with this filling the resonance voltage is determined
as follows: for a field $B\in$(4~T; 6~T): $eV_r = \hbar\omega_c$;
for $B\in$(6~T; 8~T): $eV_r = \Delta_s$ where $\Delta_s$ is the LL
spin-splitting; $eV_r = \hbar\omega_c - \Delta_s$ and for $B >
15$~T: $eV_r = 0$. These expected values are shown in
Fig.~\ref{fig:spectra} as the lengths of solid lines.

\begin{figure}
\includegraphics{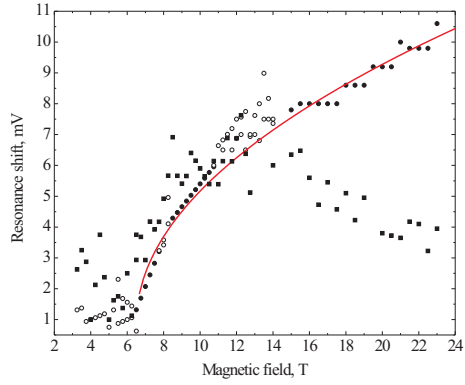} \caption{\label{fig:pseudogap} The shifts of the resonances
from its expected position versus magnetic field. The best fitted
curve to the circles is shown as solid line.}
\end{figure}

There is a remarkable shift of the experimental data from the
expected values. This shift can be associated with the Coulomb
pseudogap. Actually if one plots the difference $\Delta V_r$
between the experimental and expected values one can get data can
be described by a unified dependence all over the magnetic range
(see solid line in Fig.~\ref{fig:pseudogap}) as following one:
$\Delta V_r = \beta + \sqrt{\alpha (B - B_0)}$ where $\beta =
0.9$~mV, $\alpha = 5.2 \times 10^{-6}$~V$^2$/T, $B_0 = 6.5$~T are
the fitting parameters. It is worth noting that the value of the
LL spin-splitting $\Delta_s$ was chosen to succeed this
unified-curve description. In this case we suppose that $\Delta_s
= 0.28 \hbar \omega_c$ that corresponds to the Land\`{e} factor of
$g_* = 8.4$. This value is very close to the previous experimental
observations of the exchange-enhanced Land\`{e}
factors\cite{pseudogap}.

In Figure~\ref{fig:spectra} the experimental voltage positions of
the positive elastic feature are shown as triangles. To calculate
these positions in the single-particle model one can use
Eq.~(\ref{elastic}) setting $k = 1$. Thus the elastic feature
voltage $V_{\rm{ep}}$ can be found as follows: in the field range
$B\in$(4.5~T; 6.5~T): $eV_{\rm{ep}} = 2\hbar \omega_c$; for
$B\in$(7~T; 11~T) one can get $eV_{\rm{ep}} = 2 \hbar \omega_c -
\Delta_s$ and for $B > 13$~T one can use $eV_{\rm{ep}} = \hbar
\omega_c$. The calculated values are shown in
Figure~\ref{fig:spectra} as dashed lengths. In
Figure~\ref{fig:pseudogap}) the voltage shift $\Delta V_{ep}$ is
plotted as triangle symbols versus the magnetic field. One can see
that this elastic pseudogap shift is very close to the shift of
the coherent resonance at the magnetic field lower 12 T. However
in the magnetic field higher 12T it becomes considerably less than
the coherent shift. To understand what happens in $B = 12$~T it
would be appropriate to consider features in the negative bias. In
doing this we shall pay attention on the second feature concerning
of its amplitude. The voltage positions $V_{\rm{en}}$ of the
elastic feature are plotted in Fig.~\ref{fig:spectra} as circle
symbols.

As for the expected values they are shown with solid lengths and
calculated in account with filling factors from Eq.~\ref{elastic}
setting $k = -1$ as follows: in the field range $B \in$(4~T; 6~T)
$eV_{\rm{en}} = 0$; for $B \in$(6~T; 8.5~T) one can get
$eV_{\rm{en}} = \Delta_s - \hbar \omega_c$ and for $B > 12$~T one
can use $eV_{\rm{en}} = - \hbar \omega_c$. The
negative-elastic-feature shift has started increase in the same
field as the coherent resonance and sharply decreases at the field
range $B\in$ (8.5~T; 12~T). This sharp decrease coincides with the
second coherent resonance transfer. In this transfer the elastic
feature is very close to the second resonance that means the
cyclotron energy is close to the intersubband energy and it gets
larger in the higher field. It is interesting to note that the
shift of the positive feature takes decrease at the same field $B
= 12$~T and also in this field the first coherent resonance shifts
close to zero voltage and the tunnel spectrum becomes more
symmetric. This resonance shift is a result of the lowest LL
pinning. In this case the subband levels had very close energy in
the both 2DEGs. So we can suppose that the decrease of the
positive elastic feature takes place also when the cyclotron
energy becomes greater than intersubband energy in the both 2DEGs.

This work was supported by RFBR (grant numbers 07-02-00487,
08-02-09373-mob-z) and RAS program "Quantum nanostructures".

\end{document}